\documentclass{article}
\usepackage{graphicx}
\begin{document}
\title{Spontaneous emission of a small source of light.
\author{Jacques Moret-Bailly 
\footnote{Laboratoire de physique, Universit\'e de Bourgogne, BP 47870, F-21078 
Dijon cedex, France. email : Jacques.Moret-Bailly@u-bourgogne.fr}}}
\maketitle

{\bf The English of this paper must be corrected.}


\begin{abstract}
The usual computation of the spontaneous emission uses a mixture of classical and 
quantum postulates. A purely classical computation shows that a source of 
electromagnetic field absorbs light in the eigenmode it is able to emit. Thus in an 
excitation by an other mode, the component of this mode on the eigenmode is 
absorbed, while the remainder is scattered. This loss of energy does not apply to the 
zero point field which has its regular energy in the eigenmode, so that the zero point 
field seems more effective than the other fields for the stimulation of light emission.
\end{abstract}
\section{Introduction.}
The spontaneous emission may be considered as an amplification of the zero point 
field \cite{Weisskopf}. However it seems that the zero point field is twice more 
effective than the ordinary fields \cite{Ginzburg,Milonni,Milonni32}.

The starting point of this conclusion is the {\it classical} computation of the rate at 
which an oscillating dipole absorbs energy from a plane wave through its electric field, 
and diffracts a fraction of this energy \cite{MilonniAD}.

The radiation reaction provides the energy necessary to double the spontaneous 
excitation of a source due to the zero point field; but this computation fails because in 
the amplification of a beam of light (hot, i.e. $T > 0$ in Planck's law) the zero point 
field is not a data of the problem.

The origin of the radiation reaction is the interaction between the charges and the 
electromagnetic field they produce, leading to Abraham-Lorentz \cite{Lorentz} or 
Ford-O'Connell \cite{Ford} formula. The computations require a model of source, 
including implicit or explicit Poincar\'e stresses.

\medskip
Avoiding the choice of a model of source, using only Maxwell's equations in the 
vacuum thanks to Schwarzschild-Fokker's trick, reduces the precision of a theory, 
but allows to fix the limits of all possible computations: It will be supposed only that a 
small source made of charges radiates an electromagnetic field which obeys 
Maxwell's equations.

Section \ref{modes} recalls the theory of the modes of an electromagnetic field in the 
vacuum.

Section \ref{Fokker} shows how the Schwarzschild-Fokker's trick allows to use the 
modes with sources.

Section \ref{Source} shows that a source can absorb the whole energy of the 
eigenmode it is able to emit, while its excitation by an other mode requires a 
projection on the eigenmode which reduces the excitation.

\medskip
The following {classical} computation starts the demonstration at the beginning, 
Maxwell's equations. The electromagnetic field will be generally simply named 
\textquotedblleft field", and represented by its electric field, the magnetic part being 
implied.

\section{Mathematical modes of the electromagnetic field in the 
vacuum.}\label{modes}
In the vacuum, without sources, Helmoltz's equation for the electric field $E$, which 
may be deduced from Maxwell's equations reduces to:
\begin{equation}
{\bf \nabla^2E}-\frac{1}{c^2}\frac{\partial^2{\bf E}}{\partial t^2}=0
\end{equation}
In the whole space, or in a region limited by perfectly conducting walls, this equation 
is linear, so that its solutions named \textquotedblleft modes" build a vector space.

As there is nether sources, nor loss of energy, the energy $W$ of an electromagnetic 
field may be computed at any time by the following integration:
\begin{equation}
W=(1/2)\int[\epsilon_0{\bf E}^2+\mu_0{\bf H}^2]dV
\end{equation}
In Physics, the result must be finite, so that plane waves are necessarily 
approximations.

The energy of the sum of two modes is generally not the sum of the energies of each; 
else, the modes are orthogonal. To avoid confusions, the energy of a mode supposed 
alone in the universe will be named \textquotedblleft energy of the bare mode" or 
\textquotedblleft bare energy"; next section will show that bare energies do not exist in 
physics.

If the energy of a bare mode is finite, the mode may be normalised multiplying ${\bf 
E}$ and ${\bf H}$ by a constant such that, if $e(\nu){\rm d}\nu$ is the energy of the 
mode between $\nu$ and $\nu+{\rm d\nu}$, 
\begin{equation}
\int\frac{e(\nu){d}\nu}{h\nu}=1.
\end{equation}

\medskip
Thus orthonormal, or at least orthogonal reference frames may be built in the space of 
the modes.

Remark that introducing a new electromagnetic field into an existing electromagnetic 
field does not necessarily increase the energy; in particular, introducing a field exactly 
opposite to an existing field cancels all, so that the energy falls to zero, it is the 
absorption of the initial field. 

\section{Schwarzschild-Fokker's trick.}\label{Fokker}
These authors replace charges by convenient fields which provide the same field 
without the charges.

A source supposed invariant by a time inversion (for instance a temporary dipole) 
radiates an emission field named \textquotedblleft retardated eigenfield" of the source. 
Adding this solution of nonlinear Maxwell's equations and its Charge-Time inverse 
(absorption eigenfield or \textquotedblleft advanced field"), the charges making the 
source cancel, so that we obtain a solution of the linear Maxwell's equations. The 
advanced field is a field converging to the virtual source; used to replace a real source 
in a mathematical representation, it has generally no physical existence, while 
approximate realisations may be tried\footnote{In laser fusion experiments for 
instance.}.

Most usual sources (electron, atoms, molecules) are small in comparison with their 
distances, so that the electromagnetic field that they radiate is large close to them, 
small close to other sources. Using the trick, the absorption of the field radiated by a 
source, close to the source, requires the sum of a lot of small fields radiated by other 
sources, so that it requires a lot of time; meanwhile it remains a field. As this may be 
applied to all sources, it exists a residual field which is, at least, the field which 
remains in the dark, at 0 Kelvin. It is an absolutely normal field.

Far from sources, this field is stochastic, but it may be amplified by sources, leaving 
its pure stochastic nature\footnote{For this reason, we prefer \textquotedblleft zero 
point field" to \textquotedblleft stochastic field".}. From the building of 
electromagnetic fields by amplification of existing fields, it appears that it is physically 
meaningless to split the field $\bf E$ of a mode into a zero point part $\bf Z$ and the 
remainder: $\bf Z$ can only be obtained by an attenuation of $\bf E$ by an absorber, 
so that $\bf E-Z$ is a difference of fields existing in two different points. The flux of 
energy available in an absorber such as a photocell is proportional to the difference 
between the square of the incident field $\bf E$ and the square of the emergent field 
$\bf Z$. The first approximation which neglects the zero point field fails in the 
detection of low fields, in particular in photon counting experiments.

\medskip
The mean value of the energy of a field in a monochromatic mode is given by the 
second Planck's law:
\begin{equation}
W_\nu=h\nu\Bigl[\frac{1}{\exp(h\nu/kT)-1}+\frac{1}{2}\Bigr]
\end{equation}
It is a mean energy, subject to fluctuations.
\section{Excitation of a source: spontaneous and stimulated 
emissions.}\label{Source}
The spontaneous emission is considered as an amplification of the energy of a mode, 
an absorption as a reduction of this energy, possibly down to the zero point energy. If 
the source is a molecule whose classical eigenstates are minimums of potential, a 
transition is possible if the molecule gets enough energy to reach the threshold 
between two minimums by an absorption, at Planck's frequency. If the threshold is 
very low, as in a photocell, the fluctuations of the zero point field are sufficient to 
provide the needed energy.

\medskip
In classical electromagnetism, the absorption of an electromagnetic field is its 
cancellation by an opposite field; using Schwarzschild-Fokker's trick, a source is 
replaced by a field and the cancellation of fields may be studied at any time.

All electromagnetic fields may be decomposed on a set of orthogonal modes 
including the eigenmode of the source. All modes orthogonal to the eigenmode cannot 
exchange energy with the source (replaced by the advanced field), so that the source 
can only be excited by the component of fields on this mode.

It is physically impossible to build a \textquotedblleft spherical"\footnote{We use 
improperly \textquotedblleft spherical" for the field emitted or absorbed by a point 
source such as a dipole while it is not invariant by all rotations around an axis which 
crosses the source.} wave to excite a point source. The exciting waves are, in the 
practice, nearly plane waves. The fraction of the wave which corresponds to the 
projection of the vector representing the eigenmode of the source on the vector 
representing the plane wave is absorbed; the remainder is scattered. The 
decomposition of the field radiated by the multipolar sources into plane waves is well 
known; one of the plane waves of the decomposition subtracts from the plane 
exciting wave, the other provide scattered waves. In the general case, the scattered 
fraction is low if the source has the properties of a good antenna.

\medskip
The trick cannot be used to compute the energy, the energy must be deduced from all 
fields; this deduction is generally easy, at least at infinity, in a Fraunhofer computation: 
Set $a_iE$ ($i=0,1,\dots$) the amplitudes of the orthogonal plane waves resulting 
from the decomposition of the field radiated by a source excited by a wave of 
amplitude $E$, and $a_0$ the coefficient corresponding to the mode of the exciting 
wave. The input energy is proportional to $E^2$, the output to 
$E^2(|(1+a_0)|^2+\Sigma_{i>0}|a_i|^2)$. The scattered energy is $W_S\propto 
E^2\Sigma_{i>0}|a_i|^2$, while the total absorbed energy is $W_T\propto 
E^2(2|a_0|\cos\phi+|a_0|^2)$ where $\phi$ is the difference of the phases between the 
exciting wave and the wave scattered in the same mode; the energy effectively 
absorbed by the source, which provides its excitation is $W_E=W_T-W_S$.

The ratio $\rho=W_E/W_T$ is between 0 and 1, the last case corresponding to a well 
adapted antenna.

Studying the spontaneous emission, it is generally obtained that the zero point field is 
two times more effective than an \textquotedblleft ordinary field". The reason is that 
this ordinary field is the increase of the zero point field in a plane mode amplified by 
the exciting source. Remark that the factor 2 requires a model of source, and the 
introduction the radiation reaction; the hypothesis are the object of a lot of 
discussions \cite{MilonniAD}.

\section{Conclusion.}
The present computation of the spontaneous emission as an amplification of the zero 
point field does not apply to quantum electrodynamics which requires a free change 
of the mode of a photon through the \textquotedblleft reduction of the wave packet", 
to allow, for instance, an EPR experiment.

The field which excites a molecule, stimulating an emission, is in a mode that the 
molecule is able to absorb, that is in the eigenmode of absorption for the transition. 
The plane mode generally used to excite a molecule, must be projected on the 
eigenmode while the regular zero point field, which is in the eigenmode of absorption, 
does not require a projection.

Thus the classical theory explains that an ordinary field is less effective than the zero 
point field to stimulate an emission, while quantum electrodynamics does not.

\end{document}